%% file: ms.tex
\documentclass{article}
\usepackage{spconf,amsmath,graphicx}
\usepackage{amsthm}
\usepackage{mathtools}
\usepackage{amsfonts}
\usepackage{amssymb}
\usepackage{graphicx}
\usepackage[dvipsnames]{xcolor}
\usepackage{multirow}

\title{Improving deep learning sound events classifiers using gram matrix feature-wise correlations}
\name{Antonio Joia Neto, Andre G. C. Pacheco, Diogo Carbonera Luvizon}
\address{Samsung R\&D Institute Brazil \\ \{a.joia, andre.cp, diogo.cl\}@samsung.com}
\begin{document}
\maketitle

\input{00_abstract.tex}

\input{01_introduction.tex}
\input{03_method.tex}

\input{04_experiments.tex}
\input{05_conclusions.tex}





\bibliographystyle{IEEEbib}
\bibliography{references}

\end{document}

%% file: 00_abstract.tex
\begin{abstract}
In this paper, we propose a new Sound Event Classification (SEC) method which is inspired in recent works for out-of-distribution detection. In our method, we analyse all the activations of a generic CNN in order to produce feature representations using Gram Matrices. The similarity metrics are evaluated considering all possible classes, and the final prediction is defined as the class that minimizes the deviation with respect to the features seeing during training. The proposed approach can be applied to any CNN and our experimental evaluation of four different architectures on two datasets demonstrated that our method consistently improves the baseline models. 
\end{abstract}

%% file: 01_introduction.tex
\section{Introduction} \label{sec:intro}

The sound event classification (SEC) task consists of identifying a set of sound events in an audio recording \cite{fayek2019sound}. Designing signal processing algorithms to assess and extract this information is a key step in several applications such as multimedia indexing based on audio content, context-aware mobile devices, interactive robots, surveillance systems, among many others \cite{mesaros2018multi}. Thereby, over the past few years, the interest in SEC has been increasing and different works have been proposed to handle this task \cite{salamon2017deep, ozer2018noise, fayek2019sound, lu2020deep}.

In order to perform the SEC task, usually, the first step is to apply an algorithm to extract features from an audio sample. Common approaches are Mel-frequency cepstral coefficient (MFCC), zero-crossing rate (ZCR), and linear predictive coding (LPC) \cite{qawaqneh2017deep, lu2020deep}. Next, the extracted features can be used as inputs for a classifier, such as Support Vector Machines (SVM) \cite{pedersen2007accent}.
Recently, different works have proposed to use convolutional neural networks (CNNs) to perform audio classification \cite{lee2009unsupervised, salamon2017deep, lu2020deep}. In most of these works, the authors propose to convert the audio recordings into spectrograms, resulting in a 2D representation of frequency \textit{vs.} time \cite{lu2020deep}.
In this context, Mel-Spectogram has become a quite popular method to convert an audio signal to a 2D representation that  can be used as input for popular CNN architectures.

Traditionally in many CNNs models, the final classification is performed based on the maximum a posteriori (MAP) estimation, which does not consider the statistics of the intermediate activations directly. This architecture design can lead to an unexpected behaviour when the network's input has a different distribution from the training data. Such variations are even more common in SEC due to the intrinsic properties of audio signals, which drastically suffer from additive noise.
Differently from previous approaches, in this work we propose a methodology that directly takes into account the activations of all the network layers to produce a metric, which is used to perform classification.
This metric relies on feature-wise correlations computed using Gram Matrices across the intermediate CNN features.

The idea of extracting feature-wise correlations using Gram Matrices was previously proposed by Gatys \textit{et al.}~\cite{gatys2016image}, resulting in a breakthrough algorithm to perform image style transfer. 
Recently, Sastry and Oore \cite{sastry2020detecting} proposed the Gram-OOD, an algorithm that uses Gram Matrices to compute CNN feature-wise correlations in order to tackle the out-of-distribution (OOD) detection problem. Later, Pacheco \textit{et al.}~\cite{pacheco2020out} proposed the Gram-OOD*, a lighter version of the original algorithm that introduces a normalization step and assess fewer layers than the original method. Both algorithms are agnostic to the model and can be applied to any neural network architecture.

In this work, we propose the Gram-Classifier, an adaptation of the Gram-OOD* method that uses the Gram Matrices of the feature maps of a CNN to predict a class label for a given sample. The proposed method benefits from a statistical analysis of intermediate CNN activations, resulting in better robustness against variations.
We evaluated the proposed method for sound event classification in two datasets using four different well-known CNN architectures. The obtained results show that our method performs better than the baselines using fully-connected layers and MAP classification, resulting in an average improvement of 3\% in terms of classification accuracy. This improvement indicates that our method is able to better use the potential of feature maps generated by CNN models.

The remainder of this paper is organized as follows. In Section 2, we describe the Gram Matrix, how to compute feature-wise correlation, and the proposed methodology. In Section 3, we present the experimental evaluation and, in Section 4, we draw our conclusions.

%% file: 03_method.tex
\section{Background and Proposed Method} \label{sec:methods}

As previously mentioned, our method is inspired by \cite{sastry2020detecting} and \cite{pacheco2020out}. Therefore, in this section, we revisit the idea of features correlation using Gram Matrix and then we describe the proposed Gram-Classifier for sound event classification.

\subsection{Features correlation with Gram Matrix}
Let us consider a set of vectors $V = \{ \mathbf{v}_1, \cdots, \mathbf{v}_k \}$ in which $\mathbf{v}_k \in \mathbb{R}^z$. Essentially, $V$ is defined by the following matrix:

\begin{equation} \label{eq:matrix_of_vs}
V = 
\begin{bmatrix}
\mathbf{v}_1\\ 
\vdots\\ 
\mathbf{v}_k
\end{bmatrix} =
\begin{bmatrix}
v_{11} & \cdots & v_{1z}\\ 
\vdots & \ddots & \cdots \\ 
v_{k1} & \cdots & v_{kz}
\end{bmatrix}.
\end{equation}

\noindent The matrix composed of the vector-wise scalar product $\left \langle \mathbf{v}_i, \mathbf{v}_j \right \rangle$ is named as the Gram Matrix of $V$ \cite{boyd2018introduction}:

\begin{equation} \label{eq:G}
G = \begin{bmatrix}
\left \langle \mathbf{v}_1, \mathbf{v}_1 \right \rangle & \cdots & \left \langle \mathbf{v}_1, \mathbf{v}_k \right \rangle \\
\vdots & \ddots & \vdots \\
\left \langle \mathbf{v}_k, \mathbf{v}_1 \right \rangle & \cdots & \left \langle \mathbf{v}_k, \mathbf{v}_k \right \rangle 
\end{bmatrix}.
\end{equation}

\noindent This operation can be rewritten to the matrix formulation as:

\begin{equation} \label{eq:original_gram_mat}
    G = V^T V
\end{equation}

\noindent The scalar product between two vectors may be interpreted as a similarity measure. In other words, it expresses the vectors' correlation. In this sense, the Gram Matrix measures the pairwise correlation of the set of vectors in $V$.

\subsection{Using Gram Matrices to extract features deviation of CNN layers} 
Let us consider a trained CNN composed of $L$ activation layers in which the representation at the $l^{th}$ layer consists of $K$ feature maps, each of size $m \times n$. Considering the $l^{th}$ layer, we interpret each feature map as a vector $\mathbf{v}_k \in \mathbb{R}^{m*n}$ and stack them as a two-dimensional matrix $F_l$ similar to the one presented in Eq.~\ref{eq:matrix_of_vs}. Essentially, this matrix stores all feature maps extracted by a CNN for a given layer. Finally, as we are dealing with a classification problem, let us also consider a dataset composed of $c \in \{1, \cdots, C\}$ classes and a training ($\textrm{Tr}$), a validation ($\textrm{Va}$), and a testing ($\textrm{Te}$) partitions.

\subsubsection{Gram matrix of feature maps} \label{sec:gram_mat_feat}
The first step of the method is to compute the Gram Matrix of each $F_l$ according to Eq. \ref{eq:original_gram_mat} \cite{sastry2020detecting, pacheco2020out}:

\begin{equation} \label{eq:gram_mat}
    G_l = F_l{F_l}^T
\end{equation}

\noindent As described in Eq. \ref{eq:G}, the $k^{th}$ row of $G_l$ matrix represents the pairwise correlation between the feature map $k$ with all others. Assessing every single correlation is redundant and impracticable. Thereby, we aggregate the rows of $G_l$ to achieve the accumulated pairwise correlation for each feature map:

\begin{equation} \label{eq:accum_corr}
    \hat{g}_{lk} = \sum_{i=1}^{m*n} \left \langle \mathbf{v}_k, \mathbf{v}_i \right \rangle
    \;\; \Rightarrow  \;\;
    \hat{G}_{l} = \begin{bmatrix}
    \hat{g}_{l1} \\
    \vdots \\
    \hat{g}_{lK}
    \end{bmatrix}
\end{equation}

\noindent Lastly, in order to ensure that all $\hat{G}_l$ matrices have the same scale, we normalize its values according to \cite{pacheco2020out}:

\begin{equation} \label{eq:norm_G}
    \tilde{G}_{l} = \frac{ \hat{G}_{l} - \min(\hat{G}_{l}) }{\max(\hat{G}_{l})-\min(\hat{G}_{l} )}.
\end{equation}

\subsection{Deviation from features correlation}
Let us suppose we have a test sample $\breve{\mathbf{x}}$ and we want to compute a metric of how much the feature maps extracted from $\breve{\mathbf{x}}$ deviates from the training samples.
For this, we use the matrix of accumulated correlations $\tilde{G}_{l}$, which represents a global descriptor for each  CNN layer. This process is detailed as follows.

First, considering the training partition $\textrm{Tr}$, we compute the minimum ($\lambda$) and maximum ($\Lambda$) values in $\tilde{G}_l$ with respect to the layer $l$:

\begin{equation} \label{eq:mins}
    \lambda_{l} = \min\left[\tilde{G}_l(X)\right]
\end{equation}

\begin{equation} \label{eq:maxs}
    \Lambda_{l} = \max\left[\tilde{G}_l(X)\right]
\end{equation}

\noindent where $X$ represents all samples in the $\textrm{Tr}$. Essentially, the method assumes that $\tilde{G}_l$ may be approximated by a uniform distribution and \{$\lambda_{l}, \Lambda_{l} \}$ map the limits of this distribution given for each layer. It is important to note that this step is performed offline, i.e., we compute \{$\lambda_{l}, \Lambda_{l} \}$ and store it to be used during inference.

The next step is to generate the deviation metric, which is computed based on the following equation \cite{sastry2020detecting}:

\begin{equation}
 \delta(\lambda, \Lambda, g) = 
\begin{cases}
0 & \textrm{if} \; \lambda  \leq g \leq \Lambda \\ 
\frac{\lambda - g}{\mid \lambda \mid} & \textrm{if} \; g < \lambda \\
\frac{g - \Lambda}{\mid \Lambda \mid} & \textrm{if} \; g > \Lambda,
\end{cases}
\end{equation}

\noindent where $g$ is a single value of $G$ and $\lambda$ and $\Lambda$ are the minimum and maximum values extracted from $G$. Therefore, for the testing sample $\breve{\mathbf{x}}$, we compute $\delta$ for each layer $l$:

\begin{equation}
    \delta_l(\breve{\mathbf{x}}) = \sum_{k=1}^{K} \delta(\lambda_{l}[k],\Lambda_{l}[k],\hat{G}_{l}(\breve{\mathbf{x}})[k]).
\end{equation}

\noindent Finally, we aggregate the deviation for all layers to produce the total deviation:

\begin{equation} \label{eq:Delta}
    \Delta(\breve{\mathbf{x}}) = \sum_{l=1}^L \frac{\delta_l(\breve{\mathbf{x}})}{ \mathbb{E}_{\textrm{Va}} \left[ \delta_l \right ] }
\end{equation}

\noindent where $\mathbb{E}_{\textrm{Va}} \left[ \delta_l \right ]$ is the expected deviation at layer $l$ computed using the validation partition $\textrm{Va}$. Computing the normalized sum of layer-wise deviations helps to account for variations in the scale of layerwise deviations ($\delta_l$), which depends on the number of channels in the layer ($K$), number of pixels per channel ($m\times n$) and semantic information contained in the layer \cite{pacheco2020out}.







\subsection{Gram-Classifier: a method to use feature-wise correlations to perform classification}

Now that we presented how to compute the total deviation from feature map correlations, let us introduce our method to perform classification. We start our method by selecting the CNN layers to account for the total deviation $\Delta$. Essentially, taking into account all layer deviations $\delta$ is not efficient and may not contribute to classify a given sample. Thereby, we propose a way to select the layers of interest. As we are handling a classification problem composed of $c \in C$ classes, let us consider $\text{Tr}_c^+$ as the training samples of class $c$, and $\text{Tr}_c^-$ as the training samples of the other classes. We define the $D^{c^-}_{\delta_l}$ and $D^{c^+}_{\delta_l}$ as the data distribution of the deviation $\delta$ from the same layer $l$, i.e., the values of $\delta$ stratified per layer and class. Next, we compute the distance between both distributions $D^{c^+}_{\delta_l}$ and  $D^{c^-}_{\delta_l}$ using the Wasserstein distance ($W_d$) \cite{villani2009wasserstein} to generate a score $I_l$ for each layer $l$:

\begin{equation} \label{eq:wasserstain}
    I_l = W_d ( D^{c^+}_{\delta_l},D^{c^-}_{\delta_l} )
\end{equation}

\noindent Finally, we select the set of interest layers $l$ that have the highest values in $I_l$.

In order to perform the classification for a new test sample $\breve{\mathbf{x}}$, we compute the total deviation $\Delta(\breve{\mathbf{x}})$ -- with respect to each class and using the layers selected in the previous step --  as follows:

\begin{equation}
    \boldsymbol{\Delta}(\breve{\mathbf{x}}) = \left [ \Delta_1 (\breve{\mathbf{x}}), \Delta_2 (\breve{\mathbf{x}}), \cdots, \Delta_C(\breve{\mathbf{x}})  \right ]
\end{equation}

\noindent Each $\Delta_c$ represents how much the test sample $\breve{\mathbf{x}}$ deviates from the training feature maps with respect to the class $c$. In other words, the lower the deviation, the closer the sample is to the class $c$. Therefore, the prediction class for $\breve{\mathbf{x}}$ is determined as follows:

\begin{equation}
    \text{pred}(\breve{\mathbf{x}}) = \text{argmin} \left [ \boldsymbol{\Delta}(\breve{\mathbf{x}}) \right ]
\end{equation}

\noindent Essentially, as the total deviation over the classes relies on the correlations extracted from the Gram Matrix, we chose the class that is more similar to the test sample.

%% file: 04_experiments.tex
\section{Experimental evaluation}
\label{sec:experiments}

In this section, we carry out experiments\footnote{Code is available on https://github.com/a-joia/Gram-Classifier} in order to evaluate the performance of the proposed method on a set of trained classifier.
We use four different CNN architectures trained on two sound event classification datasets. In this section we describe the experimental setup and present the obtained results.

\subsection{Datasets and metrics}

We evaluate our method on two datasets: 

\begin{itemize}
    
    \item DCASE 2020 Task 1B\footnote{http://dcase.community/challenge2020/task-acoustic-scene-classification\#subtask-b}: an acoustic scene classification dataset composed of 40 hours of data and three major classes: indoor, outdoor, and transportation.
    
    \item DCASE 2019 Task 1A\footnote{http://dcase.community/challenge2019/task-acoustic-scene-classification-results-a}: an acoustic scene classification dataset composed of 40 hours of data and 10 different classes. The data is split into segments 10 seconds per sample.


\end{itemize}

Both DCASE 2020 and DCASE 2019 datasets have a standard training, validation, and testing partitions, however, the ground truth for the testing partition is not public available.
Thereby, we evaluated the performance of our approach using the public evaluation set. Particularly for DCASE 2019, we submitted our results on the leaderboard partition to be evaluated on Kaggle\footnote{https://www.kaggle.com/c/dcase2019-task1a-leaderboard/}. For the evaluation metric, we reported the average classification accuracy (ACC), balanced accuracy (BA).


\subsection{Experimental setup}

In order to assess the performance of the proposed method, we train four well-known CNN architectures with random weight initialization. The baseline architectures are ResNet50 \cite{he2016deep}, DenseNet-121, \cite{huang2017densely}, VGGNet-16 \cite{simonyan2014very}, and  MobileNet-v2 \cite{sandler2018mobilenetv2}. We trained all networks for 100 epochs without data augmentation and used RAdam optimizer~\cite{liu2019radam} with learning rate equals to 0.001 and batch size equals to 32. We select the model weights based on the best validation score.

For both datasets, we resampled all the segments to a rate of 16 kHz and use the 128-dimensional Mel-spectrograms as input features to each CNN. The spectrograms were extracted using 1024 bins of fast Fourier transform applied on temporal windows of 40 milliseconds with 20 milliseconds of overlap.

For each CNN model, we compared the results of the original architecture using the MAP estimation with the proposed Gram-Classifier. We also compared all the results with the reference baseline available scores for each benchmark previously described.

\subsection{Results}
In this section, we present the results obtained for the previously described datasets. In Table \ref{tab:dcase2020_results} is presented the obtained results for DCASE 2020 dataset in terms of accuracy and balanced accuracy.

\begin{table} [!htp]
\tiny
\resizebox{\columnwidth}{!}{%
\begin{tabular}{c|cc|cc}
\hline
\multirow{3}{*}{\textbf{Model}} & 
\multicolumn{4}{c}{\textbf{DCASE 2020 Task 1b}}  \\ \cline{2-5}

&\multicolumn{2}{c|}{Baseline} & 
\multicolumn{2}{c}{Gram-Classifier}   \\ 
\cline{2-5}
& ACC & BA & ACC & BA   \\ \hline

ResNet50 & 
91.54 & 91.54  & \textbf{93.66} & \textbf{93.62}  
\\


DenseNet-121 & 
93.26 & 93.35 & \textbf{93.50} & \textbf{93.45} 
\\

VGGNet-16 & 
90.36 & 90.62 & \textbf{92.38} & \textbf{92.41} 
\\


MobileNet-v2 & 
92.68 & 92.72 & \textbf{93.70} & \textbf{93.61} 
\\ 
\hline
\hline

AVG & 
91.96 & 92.05 & 93.31 & 93.27

\\
\hline
\hline
{DCASE Baseline} & 
88.00 & NA  & NA &
NA 
\\\hline 
\end{tabular}
}
\caption{Experimental results of the proposed method compared with the baseline models and the competition baseline for DCASE 2020 dataset. AVG is the average performance considering all models.}
\label{tab:dcase2020_results}
\end{table}

As we can see, the proposed method generally performs better than the CNN models using maximum a posteriori estimation and the reference baseline. Quantitatively, it provides an average improvement of around 1.35\% in the classification accuracy when comparing to all models. Comparing to the baseline, the CNN models with MAP improves the classification accuracy in around 4\% and the Gram-Classifier in around 5\%.

In Table \ref{tab:dcase2019_results}, we present the results for DCASE 2019 dataset. For this dataset, beyond comparing the CNN models performance for the public test partition, we also include the private results -- P.ACC metric in the table. 

\begin{table}[!htp]

\resizebox{\columnwidth}{!}{%
\begin{tabular}{c|cc|c|cc|c}
\hline
\multirow{3}{*}{\textbf{Model}} & 
\multicolumn{6}{c}{\textbf{DCASE 2019 Task 1A}}  \\ \cline{2-7}

&\multicolumn{3}{c|}{Baseline} & 
\multicolumn{3}{c}{Gram-Classifier}  \\ 
\cline{2-7}
& ACC & BA &  P. ACC & ACC  & BA & P. ACC  \\ \hline

ResNet50 & 
62.16  & 62.34  & 64.66  & 
\textbf{63.15} & \textbf{63.56} & \textbf{64.83}
\\

DenseNet-121 & 
63.48 & 63.61 & 64.16 & \textbf{65.22} &
\textbf{65.35} & \textbf{66.83}  
\\


VGGNet-16 & 
60.02 & 60.25 & 63.17 & \textbf{61.18} &
\textbf{61.42} & \textbf{66.83}
\\


MobileNet-v2 & 
60.31 & 60.41 & 61.83 & \textbf{62.78} &
\textbf{62.96} & \textbf{63.66} 
\\ 

\hline
\hline
AVG & 61.45 & 61.615 & 63.66 & \textbf{64.47} & \textbf{63.72} & \textbf{65.79} \\
\hline
\hline
DCASE Baseline & 
62.5 & NA & 63.00 & NA &
NA & NA 
\\\hline 

\end{tabular}
}
\caption{Experimental results of the proposed method compared to baseline models the competition baseline for DCASE 2019 dataset. AVG is the average performance considering all models.}
\label{tab:dcase2019_results}
\end{table}

Observing Table \ref{tab:dcase2019_results}, we notice that the proposed classifier method provides a better performance for all CNN models. It improves the average accuracy in around 3\% for the public test partition and in around 2\% for the private one. Considering the reference baseline, the Gram-Classifier improves the accuracy by over 2\% for both partitions.

\subsection{Discussion}

The experiment results achieved in this section indicate that using a classifier that takes into account more layers within the CNN may improve the classification performance. Intuitively, lower level feature maps store representations that may be exploited by a classifier. The main contribution of the proposed method is to provide a way to compute a metric that consider all layers across the network. However, it is worth noticing that we are using a pre-trained model, and the proposed method does not contribute during the training phase of the network. 

The results achieved to sound event classification suggest that this method is particularly appropriate for this task. The feature maps generated by convolutional networks trained on spectograms of acoustics sound events bring relevant spatial information that may be useful to improve classification. Essentially, the feature-wise correlation computed using Gram Matrix seems to be a proper approach to correlate this spatial information among the sound samples within the same class.







%% file: 05_conclusions.tex
\section{Conclusions}
\label{sec:conclusions}

In this paper, we propose a new approach to perform sound event classification using Convolutional Neural Networks (CNNs) and Gram Matrices. As described, our method computes the Gram Matrices of the features maps within the CNN and use them to determine a deviation metric, which is used to perform the classification. An advantage of this method is that it is agnostic to the CNN model, i.e., it can be easily applied to any type of model. We performed experiments using four well-known CNN architectures trained on two benchmarks. The obtained results show that our method improved the classification performance for all CNN models in both benchmarks. Quantitatively, it provides an average improvement of around 1\% to 3\% in classification accuracy. Despite the promising results, there is still room for improvement. In the near future, we intend to investigate methods to improve the representation of the features aiming to get better spatial separation.